\documentclass[aps,amsmath,amssymb,twocolumn]{revtex4}

\usepackage[parfill]{parskip}    
\usepackage{graphicx}
\usepackage{amssymb}
\usepackage{epstopdf}
\usepackage{color}
\usepackage{graphicx}
\usepackage{pdfpages}
\usepackage{soul}
\usepackage{amsmath}

\begin{document}

\title{Lagrangian formulation for an extended cosmological equation-of-state}
\author{
Grigoris Panotopoulos  {${}^{a}$
\footnote{grigorios.panotopoulos@tecnico.ulisboa.pt}
}
Il\'idio Lopes  {${}^{a}$
\footnote{ilidio.lopes@tecnico.ulisboa.pt}
}
\'Angel Rinc\'on {${}^{b}$
\footnote{angel.rincon@pucv.cl}
} 
}
\address{
${}^a$ Centro de Astrof{\'i}sica e Gravita{\c c}{\~a}o, Instituto Superior T{\'e}cnico-IST, Universidade de Lisboa-UL, 
Av. Rovisco Pais, 1049-001 Lisboa, Portugal.   
\\
${}^b$ Instituto de F\'isica, Pontificia Universidad Cat\'olica de Valpara\'iso, Avenida Brasil 2950, Casilla 4059, 
Valpara\'iso, Chile.
}

\begin{abstract}
We show that the extended cosmological equation-of-state developed starting from a Chaplygin equation-of-state, recently applied to stellar modeling, is a viable dark energy model consistent with standard scalar potentials. Moreover we find a Lagrangian formulation based on a canonical scalar field with the appropriate self-interaction potential. Finally, we fit the scalar potential obtained numerically with concrete functions well studied in the literature. Our results may be of interest to model builders and particle physicists. 
\end{abstract}

\maketitle

\section{Introduction}

In the end of the 90's the most dramatic discovery in Cosmology was the current cosmic acceleration \cite{SN1,SN2}. 
Current and more recent observational data coming from several sides provide strong evidence that we live in a spatially flat Universe expanding at an accelerating rate \cite{turner}. The origin and nature of dark energy, the component that currently dominates the evolution of the Universe and drives its acceleration, is still uncertain, and it comprises 
one of the biggest mysteries and challenges in modern theoretical Cosmology.

\smallskip 

It is easy to verify that Einstein's General Relativity \cite{GR} with radiation and non-relativistic matter only cannot lead to accelerating solutions. A positive cosmological constant \cite{einstein} on the other hand is the simplest, most economical model in a very good agreement with a great deal of current observational data. Since, however, it suffers from the cosmological constant (CC) problems \cite{zeldovich,weinberg}, other alternatives with an evolving equation of state have been studied in the literature over the years. Although some progress has been made up to now, see e.g. \cite{Garriga:2000cv,Padmanabhan:2013hqa,Mikovic:2014opa,Canales:2018tbn}, the origin of the CC problem still remains a mystery.

\smallskip

Moreover, there is nowadays a tension regarding the value of the Hubble constant, $H_0$, between high red-shift CMB 
data and low red-shift data, see e.g. \cite{tension,tension1,tension2,tension3}. The value of the Hubble constant 
extracted by the PLANCK Collaboration \cite{planck1,planck2}, $H_0 = (67-68)~\text{km/(Mpc  sec)}$, is found to be 
lower than the value obtained by local measurements, $H_0 = (73-74)~\text{km/(Mpc sec)}$ \cite{hubble,recent}. This 
tension might call for new physics \cite{newphysics}. What is more, regarding large scale structure formation data, 
the growth rate from red-shift space distortion measurements has been found to be lower than expected from PLANCK \cite{eriksen,basilakos}. Those tensions have been addressed in cosmological models with a varying cosmological constant, see e.g. \cite{Sola1,Sola2,Sola3,Sola4} and references therein. The effective cosmological equations in $\Lambda$-varying scenarios can be computed without additional assumptions, they are a natural generalization of the more standard concordance model, and they offer a richer phenomenology compared to the $\Lambda$-CDM model \cite{Oztas:2018jsu}. 
There are also some works on cosmological models with a variable Newton's constant, see e.g. \cite{VarG1,VarG2,VarG3}.

\smallskip

Regarding the aforementioned problems and possible alternatives to the $\Lambda$-CDM model, one has quite generically two choices. Either a modified theory of gravity must be assumed (geometrical DE), providing correction terms to GR on cosmological scales, or a new dynamical field (dynamical DE) with an equation-of-state (EOS) parameter $w < -1/3$ must be introduced. In the former class of models one finds for instance $f(R)$ theories of gravity \cite{mod1,mod2,HS,starobinsky} (see also \cite{NewRef1,NewRef2} for a more general class of theories, $f(R,L)$, that allow for more general couplings between matter and curvature), brane models \cite{langlois,maartens,dgp} and scalar-tensor theories of gravity \cite{BD1,BD2,leandros,PR}. In the second class one finds models such as quintessence \cite{DE1}, phantom \cite{DE2}, quintom \cite{DE3}, tachyonic \cite{DE4} or k-essence \cite{DE5}. For an excellent review on the dynamics of dark energy see e.g. \cite{copeland}. 

\smallskip

In phenomenological descriptions of dark energy models the simplest approach is to assume a concrete parameterization 
$w(z)$, where the equation-of-state parameter (EoS), $w$, is a certain function of the red-shift $z$. Another approach is to assume for the fluid component that accelerates the Universe a suitable EoS, namely a certain 
expression $f(p,\rho)=0$ relating the pressure $p$ with the energy density $\rho$. In both cases our favorite model 
is characterized by a few free parameters to be determined upon comparison with current observational data. We feel 
that the second possibility is more natural, which is the reason why we shall adopt this approach in the discussion to follow. 

\smallskip

Of particular interest is a generalized gas model \cite{CG1,CG2}, which unifies non-relativistic matter with the CC introducing a single fluid with an EoS $p=-B^2/\rho^\omega$, where $B,\omega$
are positive constant parameters, and $\omega$ takes values in the range $0 < \omega \leq 1$. Recently, an extended Chaplygin equation-of-state (GC) of the form $p=-B^2/\rho + A^2 \rho$, where a barotropic term is added to the standard Chaphygin equation-of-state with $\omega=1$, was considered in stellar modelling \cite{angel,PRL}. See also \cite{ExtCh1,ExtCh2} for earlier works on the extended Chaplygin EoS, where a more general expression for the equation-of-state may be found. The study of the extended Chaplygin model in Cosmology is further motivated by \cite{Extra1,Extra2,Extra3}, where observational data were used to demonstrate the advantage of the model.

\smallskip

Although it is convenient to study dark energy parameterizations, since for a given $w(a)$, with $a$ being the scale factor, the expansion history of the Universe is known, a more fundamental description is often needed, based on a canonical scalar field for example. In the present work our goal in two-fold: First, we propose to study this generalized equation-of-state in a cosmological context, which to the best of our knowledge is still missing, and thus filling a gap in the literature, and after that to find a Lagrangian formulation of the model based on a real scalar field minimally coupled to gravity with the appropriate self-interaction potential. 

\smallskip

Our work is organized as follows:  After this Introduction, we present the cosmological model in the next section, and the corresponding Lagrangian formulation in section 3. Finally, we summarize our work with some concluding remarks in Section \ref{Conclusions}. We work in natural units where $c=1=\hbar$.


\begin{figure*} [ht!]
	\centering
	\includegraphics[width=0.48\textwidth]{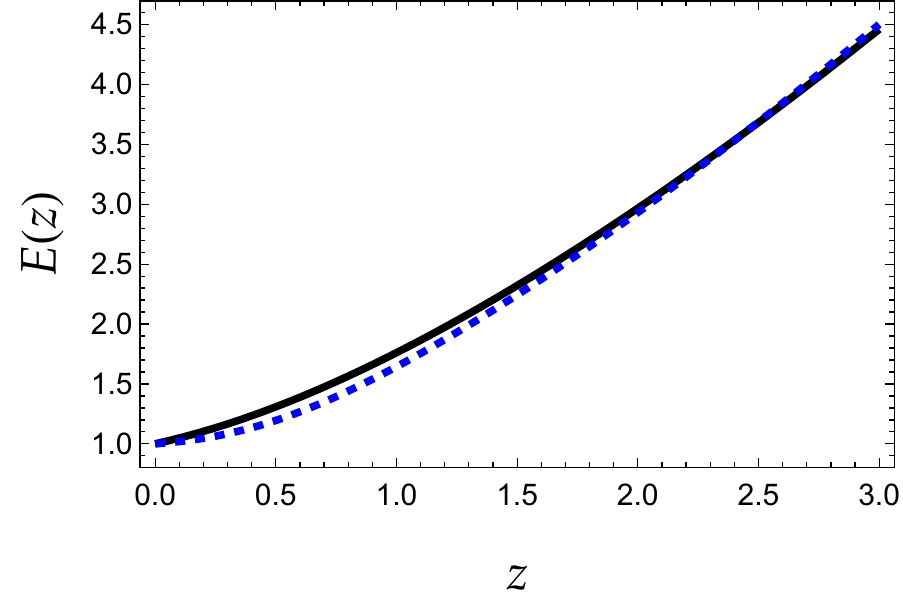} 
	\includegraphics[width=0.48\textwidth]{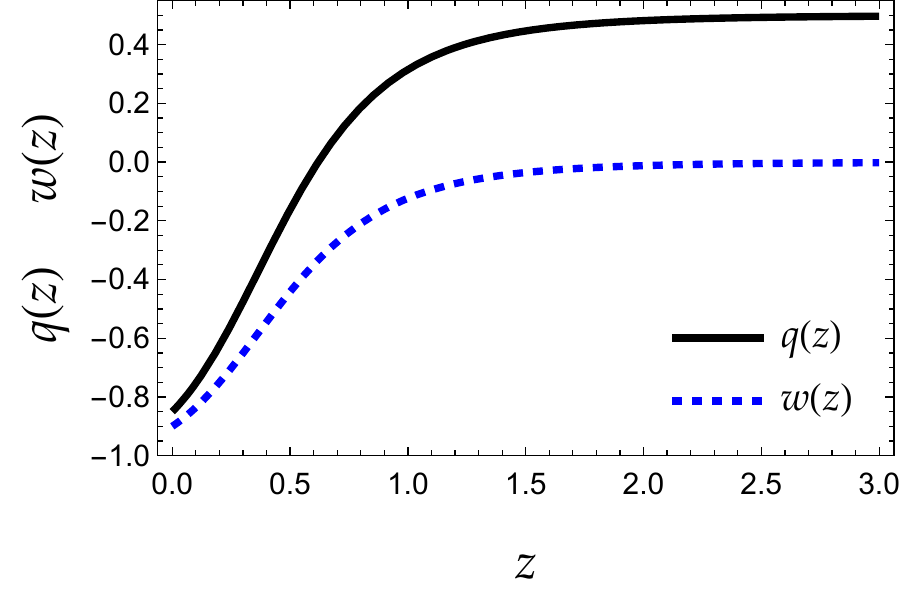} 
	\caption{
		{\bf LEFT:} $H(z)/H_0$ versus red-shift for $\Lambda$CDM model and for the extended Chaplygin
		EoS assuming $A=0.01$ and $\Omega=0.1$.
		{\bf RIGHT:} Deceleration parameter, $q,$ and effective EoS parameter, $w$, versus red-shift for 
		the extended Chaplygin EoS considered here.}
	\label{fig:1} 	
\end{figure*}


\begin{figure} [ht!]
	\centering
	\includegraphics[width=0.48\textwidth]{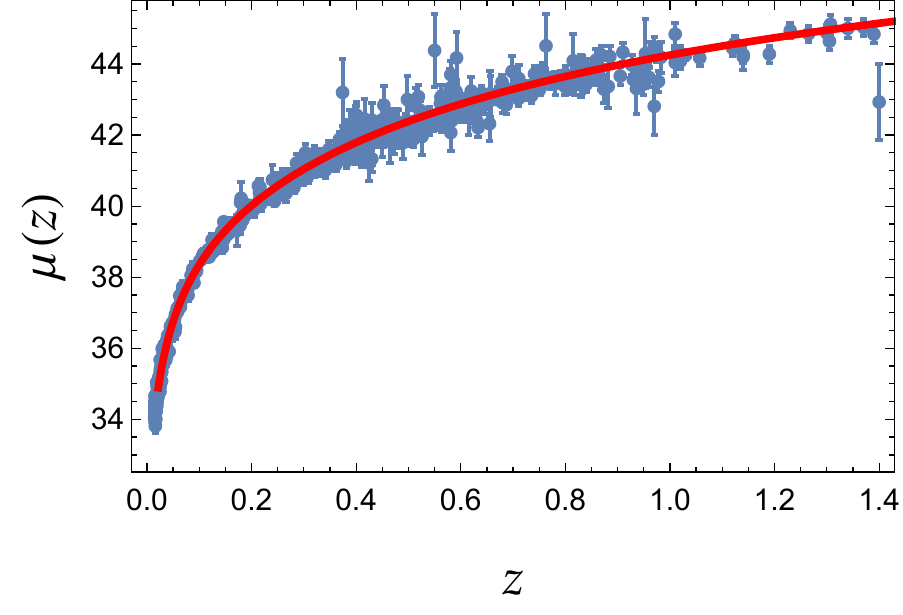}  
	\caption{
		Distance modulus versus red-shift for the Union 2 SN data as well as for
		the extended Chaplygin EoS considered here. 
		}
	\label{fig:2} 	
\end{figure}


\section{Current cosmic acceleration}

Assuming for the metric tensor a flat FRW Universe
\begin{equation}
ds^2 = - dt^2 + a(t)^2 \delta_{ij} dx^i dx^j
\end{equation}
with $t$ being the cosmic time and $a(t)$ being the scale factor, and a perfect fluid for the matter content of the Universe
\begin{equation}
T_\nu^\mu = \text{diag}(-\rho, p, p, p)
\end{equation}
with $\rho$ being the energy density and $p$ being the pressure of the fluid, Einstein's field equations give rise to the two Friedmann equations
\begin{eqnarray}
H^2 & = & \frac{8 \pi G \rho}{3} \\
\dot{H} & = & - 4 \pi G (\rho + p)
\end{eqnarray}
where $G$ is Newton's constant, $H \equiv \dot{a}/a$ is the Hubble parameter, and an overdot denotes differentiation with respect to cosmic time. Those two equations imply conservation of energy
\begin{equation}
\dot{\rho}+3 H (\rho + p) = 0
\end{equation}

\smallskip

The above cosmological equations can be combined to compute the equation-of-state 
parameter, $w$, and the deceleration parameter, $q$, defined by
\begin{eqnarray}
w & \equiv & \frac{p}{\rho} \\
q & \equiv & - \frac{\ddot{a}}{a H^2}
\end{eqnarray}
Introducing the red-shift, $1+z = a_0 / a$, as well as the dimensionless function $E(z) = H(z) / H_0$, with $H_0$ and $a_0$ being the present values of the Hubble parameter and the scale factor, respectively,
$q$ and $w$ are found to be
\begin{eqnarray}
w(z) & = & -1 + \frac{2}{3} (1+z) \frac{E'(z)}{E(z)} \\
q(z) & = & -1 + (1+z) \frac{E'(z)}{E(z)}
\end{eqnarray}
as functions of the red-shift, where the prime denotes differentiation with respect to red-shift.

\smallskip

In the present work we shall consider the extended EoS
\begin{equation}
p = -\frac{B^2}{\rho} + A^2 \rho
\label{eq:eos}
\end{equation}
characterized by two free parameters $A,B$, out of which $A$ is dimensionless, while $B$ has dimensions of pressure and energy density. Conservation of energy may be directly integrated to obtain $\rho$ as a function of red-shift
\begin{equation}
\rho(z) = \left( \frac{B^2 + \rho_1^2 (1+z)^{6 (1+A^2)}}{1+A^2} \right)^{1/2}
\end{equation}
where $\rho_1$ is an integration constant, and therefore the dimensionless Hubble parameter is found to be
\begin{equation}
\frac{H(z)}{H_0} = E(z) = \left( 1-\Omega + \Omega (1+z)^{6 (1+A^2)} \right)^{1/4}
\label{eq:Ez}
\end{equation}
where we have introduced $\Omega$ instead of $B$, with $\Omega$ defined by
\begin{equation}
\Omega \equiv 1 - \left( \frac{8 \pi}{3} \right)^2 \frac{B^2}{H_0^4 m_{pl}^4 (1+A^2)}
\end{equation}
with $m_{pl}=1.22 \times 10^{19} GeV$ being the Planck mass. 
Notice that in the special case where $A=0$, corresponding to Chaplygin gas, the energy density takes the form
\begin{equation}
\rho(z) = \left[ B^2 + \rho_1^2 (1+z)^{6} \right]^{1/2}
\end{equation}
which in the two limiting cases takes the following form, namely at late times, $z \ll 1$, it becomes a constant
\begin{equation}
\rho(z) \approx \sqrt{B^2 + \rho_1^2}=\text{const}., \qquad z \ll 1
\end{equation}
while in the past, $z \gg 1$, it takes the form
\begin{equation}
\rho(z) \approx \rho_1 (1+z)^3, \qquad z \gg 1
\end{equation}
which is the expression for the energy density of non-relativistic matter dominating the expansion of the Universe in the matter era. 

\smallskip

A viable model may be obtained if the expansion history is similar to the one corresponding to $\Lambda$-CDM model. Thus, to exemplify that our model is compatible with current observations, in the following we shall consider the case where $A=0.01$ and $\Omega=0.1$. In terms of $A,\Omega$, the $B$ parameter is computed to be
\begin{eqnarray}
B & = & \sqrt{1+A^2} \: \sqrt{1-\Omega} \: \rho_{cr} \\
\rho_{cr} & \equiv & \frac{3H_0^2}{8 \pi G}
\end{eqnarray}
and it is found to be $B = 0.95 \rho_{cr}$ for $A=0.01$ and $\Omega=0.1$. The expansion history, $E(z)$, and the 
effective equation-of-state, $w(z)$, together with the deceleration parameter $q(z)$ are shown in two panels of Fig.~\eqref{fig:1}. Also, the distance modulus $\mu(z)$ is shown in Fig.~\eqref{fig:2}. The observational data from the Union 2 compilation \cite{union} are also shown for comparison reasons. Those figures clearly demonstrate that the model
is capable of accelerating the Universe in an excellent agreement with the SN data, while in the past the matter era is recovered.


\begin{figure*}[ht!]
	\centering
	\includegraphics[width=0.48\textwidth]{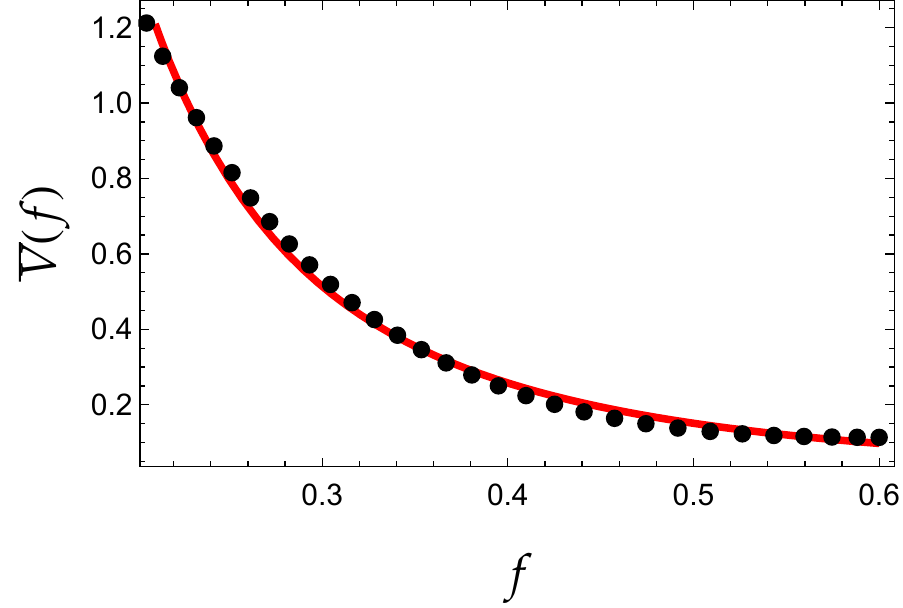} \
	\includegraphics[width=0.48\textwidth]{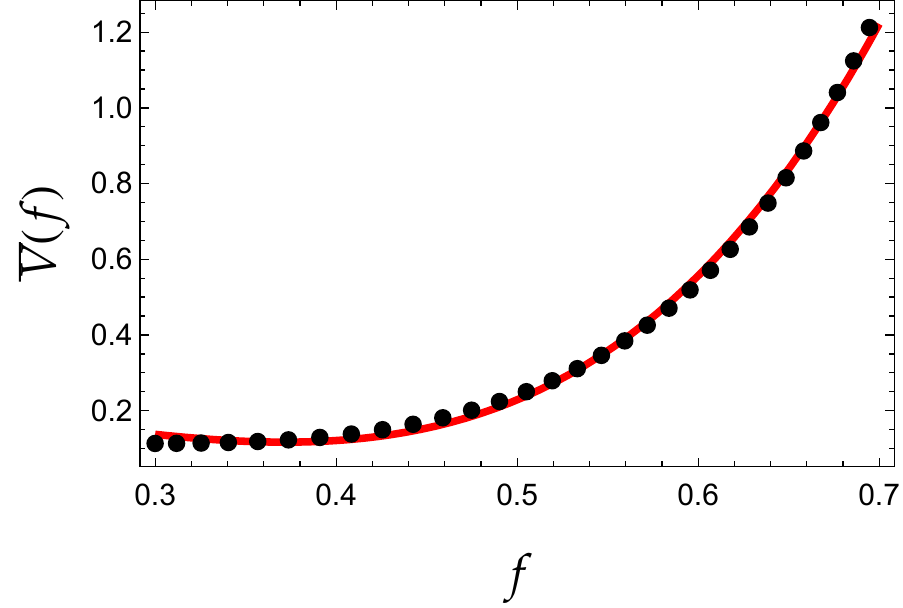} 
	\\
	\caption{
		{\bf LEFT:} Normalized scalar potential $\tilde{V} = V / (H_0 m_{pl})^2$ versus normalized scalar field $f = \phi / m_{pl} $ for the negative power-law case. The points correspond to the numerical results, while the red line corresponds to the fitting curve $\tilde{V}(f)=M^4 / f^\alpha$.
		{\bf RIGHT:} Same as left panel, but for the Higgs-like potential $\tilde{V}(f)=V_0+\bar{a} f^2+\bar{b} f^4$.
	}
	\label{fig:3} 	
\end{figure*}


\section{Lagrangian formulation}

It is always advantageous to aim at a more fundamental description, and ideally the model should be formulated starting 
from a Lagrangian. The simplest approach is probably to look for a Lagrangian formulation based on a canonical scalar field. Dealing with scalar fields is quite simple thanks to the fact that they do not carry any indices. Moreover, they arise in many different contexts in modern Particle Physics. Some well-known examples are the following: i) Higgs bosons required to break electroweak symmetry, and to give masses to particles \cite{higgs1,higgs2}, ii) pseudo-Goldstone bosons associated with explicit breaking of additional global symmetries \cite{freese}, iii) moduli coming from Superstring theory compactifications \cite{moduli1,moduli2,moduli3,moduli4,moduli5}, iv) scalar fields included in supermultiplets in supersymmetric models \cite{martin} and theories of supergravity \cite{nilles}, to mention just a few.

\smallskip

If the dark energy equation-of-state remains always in the range $-1 < w(z) < 1$, the assumed dark energy parameterization can be realized using a canonical scalar field
$\phi$ with an appropriate positive self-interaction potential $V(\phi)$. It is known that a minimally coupled scalar field behaves like a perfect fluid with
pressure and energy density given by \cite{Riotto,Mukhanov:2005sc}
\begin{align}
p_{\phi} &= \frac{1}{2} \dot{\phi}^2-V
\\
\rho_{\phi} &= \frac{1}{2} \dot{\phi}^2+V
\end{align}
Therefore, using the Friedmann equations one can determine both the potential $V(z)$ and the scalar field $\phi(z)$ as functions of red-shift as follows
\begin{align}
\tilde{V}(z) & \equiv \frac{V(z)}{H_0^2 m_{pl}^2} = \frac{E(z)^2}{8 \pi} [2 - q(z)]
\\
\tilde{\phi}'(z) &\equiv \frac{d(\phi/m_{pl})}{dz} = \pm \left( \frac{E'(z)}{4 \pi (1+z) E(z)} \right)^{1/2}
\end{align}
It is easy to verify that the previous expressions in the case of the $\Lambda$CDM model give $V(z)=\text{constant}$ and $\phi'(z)=0$. Those expressions give the scalar potential $V(\phi)$ in parametric form $\phi(z)$ and $V(z)$. The observational constraints on the previous set of equations are coming from the cosmological equation of state (equation \ref{eq:eos}) trough  $E(z)$ (equation \ref{eq:Ez}).

\smallskip

In the case where $\phi'(z) < 0$, the scalar potential may be fitted by a negative power-law of the form \cite{Ratra:1987rm}
\begin{equation}
\tilde{V}(f) = \frac{M^4}{f^\alpha}
\end{equation}
with $\alpha=2.4$ and $M=0.413$, and where $f=\phi / m_{pl}$. In the case where $\phi'(z) > 0$, the scalar potential may be fitted by a Higgs-like potential of the form \cite{SSB1,SSB2,SSB3,SSB4,SSB5}
\begin{equation}
\tilde{V}(f) = V_0 + \bar{a} f^2 + \bar{b} f^4
\end{equation}
with $V_0=0.287$, $\bar{a}=-2.464$ and $\bar{b}=8.923$. Both potentials are shown in the two panels of Fig.~\eqref{fig:3}, which shows that the model considered here not only is a viable dark energy model, but at the same time it admits a Lagrangian formulation based on scalar potentials well-studied in the literature.

\section{Conclusions}\label{Conclusions}

To summarize our work, in the present article we have studied an extended Chaplygin equation-of-state, recently 
applied to stellar modeling, in a cosmological context. First, we showed that for a suitable choice of the parameters
of the equation-of-state, it is a viable dark energy model with an expansion history very similar to the one corresponding to $\Lambda$-CDM model. In particular, as shown in Figures 1 and 2 in the text, the model is capable of accelerating the Universe in excellent agreement with supernovae data, while in the past the matter era is recovered. Next, we have obtained a Lagrangian formulation of the model based on a canonical scalar field with the appropriate self-interaction potential. Finally, we have fitted the scalar potentials obtained numerically with concrete functions well studied in the literature. We have focused here on inverse power-law as well as Higgs-like potentials, which are well-motivated and studied in older works on the inflationary Universe.


\section*{Acknowlegements}

We wish to thank the anonymous reviewer for useful comments and suggestions.
The authors G.~P. and I.~L. thank the Fun\-da\c c\~ao para a Ci\^encia e Tecnologia (FCT), Portugal, for the financial support to the Center for Astrophysics and Gravitation-CENTRA, Instituto Superior T\'ecnico, Universidade de Lisboa, through the Project No.~UIDB/00099/2020 and No.~PTDC/FIS-AST/28920/2017. 
The author A.~R. acknowledges DI-VRIEA for financial support through Proyecto Postdoctorado 2019 VRIEA-PUCV.


\end{document}